\documentclass[preprint,12pt]{aastex}

\newcommand{\etal} {et~al.}

\newcommand{\ccd}{{\small CCD}}

\newcommand{\chandra}{{\it Chandra}}

\newcommand{\epic}{{\small EPIC}}

\newcommand{\letgs}{{\small LETGS}}
\newcommand{\hrc}{{\small HRC}}
\newcommand{\hrcs}{{\small HRC-S}}

\newcommand{\mos}{{\small MOS}}

\newcommand{\mul}{{\small $\mu$~Lep}}
\newcommand{\mulb}{{\small $\mu$~Lep-B}}
\newcommand{\B}{B-star}

\newcommand{\rgs}{{\small RGS}}
\newcommand{\rosat}{{\small \it ROSAT}}

\newcommand{\uv}{{\small UV}}
\newcommand{\xmm}{{\small \it XMM-Newton}}

\newcommand{\x}{X-ray}
\newcommand{\xs}{X-rays}
\newcommand{\ms}{{\small MS}}
\newcommand{\mss}{{\small MS} star}
\newcommand{\pms}{{\small PMS}}
\newcommand{\pmss}{{\small PMS} star}

\newcommand{\cmsq}{cm$^{-2}$}
\newcommand{\kms}{km~s$^{-1}$}
\newcommand{\sm}{s$^{-1}$}

\begin{document}

\title{Resolving X-ray sources from B-stars spectroscopically:\\
The example of $\mu$ Lep}

\author{Ehud~Behar\altaffilmark{1},
Maurice~Leutenegger\altaffilmark{2}, Rami~Doron\altaffilmark{3},
Manuel~G\"udel\altaffilmark{4}, Uri~Feldman\altaffilmark{5},
Marc~Audard\altaffilmark{2}, Steven~M.~Kahn\altaffilmark{6}, }

\altaffiltext{1}{Physics Department,
  Technion, Haifa 32000,
  Israel; behar@physics.technion.ac.il}

\altaffiltext{2}{Columbia Astrophysics Laboratory,
  Columbia University,
  550 West 120th Street,
  New York, NY 10027}

\altaffiltext{3} {Weizmann Institute of Science, Rehovot 76100,
Israel}

\altaffiltext{4}{Paul Scherrer Institut, W\"urenlingen \&
Villigen, 5232 Villigen PSI, Switzerland}

\altaffiltext{5}{Artep Inc., Ellicot City, MD 21042}

\altaffiltext{6}{KIPAC, SLAC, 2575 Sand Hill Road, Menlo Park,
CA~94025}


\begin{abstract}

We present high-resolution \x\ observations of the chemically
peculiar late-type \B : \mul. However, we find spectroscopic and
astrometric evidence, which show that the \xs\ are not traced back
to the \B\ itself, but rather to a previously unresolved
companion: \mulb , whose \x\ spectrum resembles that of a
coronally active source. We discuss the possibility that \mulb\ is
a pre-main-sequence companion, most likely of the non-accreting
magnetically-active type.

\end{abstract}


\section{Introduction}

Bright X-ray emission from isolated late-type \B s (and early
A-stars) remains enigmatic, as conventional stellar theory
predicts these stars neither to have the magnetically active
coronae of cool stars nor to eject sufficiently intense stellar
winds like those of hot O- and early B- stars. Therefore, it was
surprising that a deep \rosat\ survey detected as many as 86 late
\B s (Bergh\"ofer, Schmitt, \& Cassinelli 1996). In an effort to
explain these results, \citet {hubrig01} searched for companions
to the alleged \x -bright \B s. A companion, it was presumed,
could be the actual \x\ source. In 24 cases out of a selected list
of 49 \B s, evidence for an active late-type companion was {\it
not} found, leaving the high level of \x\ flux associated with
these \B s a mystery. Until recently, the discussion in the
community on whether B-stars are intrinsic \x\ emitters or not has
suffered from the limitations on our ability to resolve the \B s
from their putative low-mass companions.

The superb angular resolution available for imaging with the
\chandra\ \x\ telescope has allowed \citet {stelzer03} to resolve
a handful of \B s, previously suspected as intrinsic \x\ emitters,
from the position of their associated \x\ sources. However, in
some cases even the angular resolution of \chandra\ is
insufficient (e.g., HD~1685~A in \citet {stelzer03}) and a
different approach is required. We have constructed and proposed a
series of \x\ spectroscopic tests, which provide a scheme to
determine whether the \xs\ emanate from the immediate vicinity of
the \B\ or not. These methods are particularly useful in the cases
where the source can not be angularly resolved by direct imaging.

Some of the \x\ brightest, presumably isolated, late-type \B s are
chemically peculiar (CP) stars, whose atmosphere is significantly
enriched with peculiar elements. Since highly-charged ions of any
heavy element emit bright lines in the \x\ band, the potential
observation of peculiar elements in the \x\ spectrum provides a
powerful tool for testing whether or not the \xs\ are due to the
\B. With that in mind, we chose to observe \mul, a late-type B9,
non-magnetic, CP HgMn \B, also known as HR~1702 or HD~33904, with
the \xmm\ and \chandra\ grating spectrometers. As detected by
\rosat\ \citep {bergh96}, \mul\ is the \x -brightest late-type
HgMn \B. Most importantly, \mul\ was included in the survey of
\citet {hubrig01} for which an \x\ active companion was searched
for, but not found.

\section{Observations}


The target \mul\ was observed with \xmm\ on 2003 March~23-24 for
47~ks. All three \epic\ cameras were operated in full-frame mode
with a thick filter and the two Reflection Grating Spectrometers
(\rgs) were operational simultaneously.
The \epic\ and \rgs\ data were processed with SAS version 5.4.1.
Standard event filtering procedures were followed including
background subtraction using off-target \ccd\ regions. \mul\ was
also observed with \chandra\ in the \letgs- \hrcs\ (Low Energy
Transmission Grating with the High Resolution Camera)
configuration on 2003 December~1-2 for 64~ks and on 2004
January~12 for 37~ks. The data were processed with CIAO version
3.0.

\subsection{Astrometry and Timing}

Using the zero-order \letgs\ image to determine the position of
the \x\ source, we obtain R.A.~= 5$^h$12$^m$55\fs 85, Dec.~=
-16\arcdeg 12\arcmin 19\farcs 91. The {\it Hipparcos} catalog
position corrected for proper motion is R.A.~= 5$^h$12$^m$55\fs
913, Dec.~= -16\arcdeg 12\arcmin 19\farcs 749 \citep {hipparcos}.
The discrepancy of 0\arcsec .93 is about a 3$\sigma$ offset from
the astrometric accuracy of \hrcs \footnote
{http://asc.harvard.edu/cal/ASPECT/celmon/}. This already hints to
the \x\ source not being associated with the \B. Nevertheless, we
proceed with the spectroscopic diagnostics below. If indeed there
is a secondary source separated by about 1\arcsec\ from \mul, it
has to be faint in the K~band as the separation achieved by \citet
{hubrig01} was as low as 0\arcsec .3 and increasingly sensitive
for larger separations. Alternatively, the separation may be
variable. At \mul 's distance of 56.5~pc, 0\arcsec .93 corresponds
to a projected distance of 52~AU.

The \epic\ \mos 1 and \letgs -zero-order light curves are shown in
Fig.~1. In the \mos\ light curve, two short duration flares
($\sim$~10~ks) are observed in which the peak flux is about 1.5
and 2.0 times what seems to be the average non-flare flux level.
\letgs\ just catches a similar flare ($\times$1.8)
declining, with subsequent flux variations of as much as 40\%
thereafter. The variability is observed on time scales of a few
hours.

\subsection{Spectroscopic Diagnostics}

The \letgs\ spectrum is of insufficient quality due to the high
background level of the \hrc. We focus therefore on the \rgs\
spectrum. \rgs\ 1 and \rgs\ 2 recorded 1400 and 1635 source
counts, respectively. The total flux measured from 6 to 37~\AA\
(0.34~- 2.1~keV) is (1.15~$\pm$ 0.03) $\times$~10~$^{-12}$~erg~\sm
~\cmsq. This can be compared with the flux of (2.4~$\pm$ 0.23)
$\times$~10~$^{-12}$~erg~\sm ~\cmsq\ measured over the 0.1~-
2.0~keV band with \rosat\ \citep {bergh96}. The discrepancy is
reasonable considering the slightly wider \rosat\ band and the
fact that the source flux varies typically by a factor of 2.
Assuming the \x\ source to be at the same distance as \mul\
(56.5~pc), the \rgs\ flux translates into an \x\ luminosity in the
\rgs\ band of $L_x$~= (4.4~$\pm$~0.1)~$\times$~10~$^{29}$~erg~\sm.

The full \rgs\ spectrum is presented in Fig.~2 along with a simple
model that serves mainly to highlight the identified bright lines.
The model is an ion-by-ion fit to the data similar to the method
used in \citet {behar01} and in \citet {brinkman01}. The spectrum
features many emission lines and little continuum. \ion{O}{8}
Ly$\alpha$ appears to be slightly (red)shifted from its rest-frame
position of 18.969~\AA. It is measured with \rgs 1 at 18.988~$\pm$
0.007~\AA\ and with \rgs 2 at 18.998~$\pm$ 0.006~\AA, i.e.,
redshifted by 300~$\pm$~110 and 460~$\pm$~90~\kms, respectively.
This could hint to rapid orbital motion of a late-type companion
or to a wind. However, the positions of other bright lines are all
formally consistent with the rest-frame wavelengths to within
200~\kms. Therefore, we prefer to interpret the \ion{O}{8}
Ly$\alpha$ anomaly as a calibration/statistics artifact. Line
widths are unresolved, providing only a not-very-constraining
upper limit of $\sigma_v~\simeq$ 500~\kms\ at 19~\AA\ for the
turbulent velocity width.

The He-like triplets are vital to the determination of the
distance of the \x\ source from the \B. In the presence of the
high \uv\ flux of the \B, intensity would be transferred from the
forbidden line ($f$) to the intercombination line ($i$) \citep
{gabriel69}. The best measured He-like triplet in the \rgs\
spectrum is that of \ion{O}{7}. The unperturbed (low-density,
no-\uv) value of the \ion{O}{7} $f/i$ intensity ratio is $\sim$
4.4, which then decreases rapidly as the distance between the \uv\
(\B) and \x\ sources is reduced.

The region of the spectrum containing the bright \ion{O}{7} lines
is shown in Fig.~3. Although the moderate statistics hamper
accurate determination of the line fluxes, the $f/i$ ratio is
obviously high and a lower limit for this ratio can be readily
obtained. The $i$ line ($\lambda$~= 21.801~\AA) can barely be
identified in Fig.~3. Indeed, the best-fit $i$ line flux is nil
with an error of 2.0~$\times$~10$^{-5}$~ph~\sm\ \cmsq. The
best-fit fluxes for the $r$ (21.602~\AA) and $f$ (22.097~\AA)
lines are 10.5~$\times$~10$^{-5}$ and
6.0~$\times$~10$^{-5}$~ph~\sm\ \cmsq, respectively, neither of
which is tightly constrained. The upper limit of
2.0~$\times$~10$^{-5}$~ph~\sm\ \cmsq\ for the $i$ line flux yields
a lower limit to the $f/i$ ratio of 3.0, although a higher value
($>$ 4.0) is much more likely. The three-line model in Fig.~3
shows the best-fit fluxes for the $r$ and $f$ lines, but the above
upper limit for the $i$ line flux.

The implication of this measurement is demonstrated in Fig.~4
where we show the theoretical $f/i$ ratio as a function of the
distance of the \x\ source from \mul\ assuming it is a perfect
blackbody at $T=$ 12,400~K \citep {adelman00}. It can be seen that
even at the conservative limit of $f/i$~= 3.0, the \x\ source has
to be at least 10.4 stellar radii ($R_*$) away from the \B. We
take this as evidence for the \xs\ not originating from the \B\
itself, but rather from a previously unresolved companion: \mulb.
With surface gravity of $\log g$~= 3.7 (cm~s$^{-2}$)
\citep{adelman00} and a mass of 3.5~$M_{\sun}$
\citep[from][]{schaller92}, 10.4~$R_*$ for \mul\ corresponds to
$\simeq$ 0.2~AU, which is much less than the estimate of
$\sim$~50~AU obtained in \S2.1. Indeed, the separation is probably
much larger than 10.4~$R_*$, only that the moderate statistics of
the spectrum call for caution. A similar result: $f/i~>$~3.0 is
found using the \ion{Ne}{9} triplet, where the $f$-line is even
stronger relatively, only the blending of the $i$-line with
\ion{Fe}{19} lines makes the exact determination of the
\ion{Ne}{9} $f/i$ ratio more model dependent.

Another powerful diagnostic tool is the potential detection of
emission lines of peculiar elements. As a HgMn star, the
abundances of elements such as Hg and Mn are enriched in the
atmosphere of \mul\ by a few orders of magnitude compared to solar
composition. If the \x\ source is on \mul, we would naively expect
to observe lines of these elements in the \x\ spectrum. The
strongest, isolated line expected is that of Ne-like \ion{Mn}{16}
at 16.62~\AA, which is the isoelectronic analog of the 15.01~\AA\
resonant line of \ion{Fe}{17}.
No clear Mn line is identified. Nonetheless, an upper limit to the
line flux and, thus, to the Mn abundance can be obtained. An
utterly prudent estimate (essentially ascribing the noise in that
range to the line) gives an upper limit of
1.0~$\times$~10$^{-6}$~ph~\sm\ \cmsq\ for the \ion{Mn}{16} line
flux, while the flux of the Fe line at 15.01~\AA\ is
1.0~$\times$~10$^{-5}$~ph~\sm\ \cmsq. Owing to the comparable
collisional excitation rates of the two lines (that of Mn is even
10\% larger), this translates into a Mn/Fe abundance ratio of
$\sim$~0.1 at the most (again, probably much less), in stark
contrast with the ratio of 1.5 observed on \mul\ \citep
{adelman00}. (The Mn/Fe solar value is 0.008.) This constitutes
more proof that the \x\ source is not on \mul. We note that strong
emission by highly ionized Hg is also expected at $\sim$~43~\AA\
\citep{doron02}, which is just outside the \rgs\ range.

In short, there is accumulating spectroscopic evidence, namely the
high $f/i$ ratios and the absence of peculiar elements, which
demonstrates that the \x\ source is not on \mul, but rather at an
appreciable distance from it. Furthermore, the \x\ source is not
due to stellar winds from the \B, nor to any other form of gas
that has undergone chemical fractionation inside the \B\
atmosphere. Wind shocks are also ruled out by the short flares and
by the observed high \x\ temperatures (see \S3).

\section{Discussion: The Nature of \mulb}

Having established that the \xs\ are not coming from \mul, the \x\
source companion \mulb\ is of interest by its own merits. The goal
of this section is to try and shed some light on the nature of
\mulb\ although the absence of optical or IR detections impedes
conclusive results. Hence, $L_{bol}$ for \mulb\ is unknown and
since the \x\ emission ($L_x$) is not due to winds nor is it
directly related to the \B, $L_x/L_{bol}$(\B) is not a meaningful
parameter. Thus, \mulb\ may be a cool main sequence (\ms), or a
pre-\ms\ (\pms) companion. \pmss s have been suggested as the \x\
companions of \B s, as the \B\ system is naturally young enough
for a low-mass companion to be in its early evolutionary stages.
Assuming $L_{bol}$~= 8.7$\times$10$^{35}$~erg~\sm\ for \mul\
\citep{bergh96}, the tables of \citet{schaller92} give an
estimated age of 19~Myr. IR detections provide evidence for \pms\
companions to \B s of similar ages \citep {stelzer03}. \pmss s are
usually also radio sources, but \mul\ has not been observed
properly in the radio. The \x\ emission mechanism in \pmss s is
thought to be magnetic \citep {feigelson99}, although shocks have
also been suggested \citep {kastner02}. Either way, it is well
established that \pmss s are bright \x\ sources.

For \mss s, there is a strong correlation between X-ray emission
and rotational velocity: $L_X$ = 10$^{27} (v \sin i)^2$~erg~\sm\
\citep {pallavicini81}. If \mul\ is a tidally locked binary, a
plausible scenario which could explain the chemical peculiarity,
its low rotational velocity: $v \sin i~<$ 10~\kms\ \citep{abt95}
would imply $L_X$ $<~10^{29}$~erg~\sm\ for a cool star companion.
This is significantly lower than observed: $L_x$~=
(4.4~$\pm$~0.1)~$\times$~10~$^{29}$~erg~\sm. On the other hand,
\pmss s have an opposite, although not as tight, correlation where
\x\ emission decreases with rotational velocity \citep[see Fig.~7
in][]{feigelson03}. Estimating the rotational period of \mul\ to
be 2$\pi R_*/v~>$~22$\sin i$ days, we find that the present $L_X$
value is consistent with the correlation of \citet{feigelson03}.
Therefore, we favor the identification of \mulb\ as a \pmss. The
\x\ luminosity and variability of \mulb\ are also consistent with
previous \x\ observations of \pmss s: The \x\ luminosity is well
within the \pms\ range of $L_x$~= 10~$^{28}$ to 10~$^{31}$
~erg~\sm\ and the variability (Fig.~1) is in line with the fact
that \x\ emission of \pmss s is dominated by flares. If indeed
\mulb\ is a \pmss, at 56.5~pc it is arguably the closest one to
earth.

The \x\ thermal and chemical structure of \mulb\ can be tested
against general characteristics of \pmss s. Although we leave a
more elaborate spectral analysis for future work, there are a few
points that can already be made here. A range of temperatures is
observed in the \rgs\ spectrum. High temperatures above 1~keV are
observed through the well identified Fe-L ions up to \ion{Fe}{24}
(see Fig.~2) and by Mg, Si, and S K-shell lines clearly detected
with the \epic\ cameras (not shown). Conversely, the 2p--3d lines
of \ion{Fe}{16} are identified at $\sim$~15.2~\AA\ (Behar et~al.
2001) and are indicative of temperatures as low as $kT_e$~=
0.2~keV. Even lower temperatures may be probed after we do a
careful job of identifying the L-shell lines of mid-Z elements,
e.g., Ca, Ar, \& S \citep {jaan03}.
Higher temperatures: $kT_e \ge$~2~keV have been suggested in \x\
measurements of \pmss s \citep{feigelson99}. There is no evidence
in our spectrum for such high temperatures. Since two other state
of the art grating observations of T Tauri stars (TTSs) \citep
{kastner02, kastner04} do not show such high temperatures either,
we suspect that temperatures as high as 2~keV may not be
ubiquitous to \pmss s.

A robust method for distinguishing between accreting and
non-accreting \pmss s by means of \x\ spectra alone has yet to be
established. Nonetheless, between the two well studied TTSs:
TW~Hydrae (accreting) and HD~98800 (non-accreting) \citep
[][respectively]{kastner02, kastner04}, the present spectrum is
more similar to that of HD~98800. The emission measure
distributions of both \mulb\ and HD~98800 are broad, the latter
covering the temperature range of 0.2~-- 1.0~keV. The strong O \&
Ne lines compared to those of Fe in the present spectrum (Fig.~2)
also resemble the ratios observed in HD~98800 and are very
different from the abundance anomaly found in TW~Hya. Finally, the
high densities (10~$^{13}$~cm$^{-3}$) observed for TW~Hya through
quenched forbidden lines are not observed for \mulb. In short,
\mulb\ is probably a non-accreting magnetically-active \pmss.

\section{Conclusions}

Using high-resolution grating spectroscopy, we have discovered the
\x\ source \mulb\ to be distinct from the \B\ \mul. This result is
based on the \x\ source being 0.\arcsec 93 from the nominal
position of the \B, but even more conclusively by two
spectroscopic findings: 1. The observed forbidden lines of the
He-like triplets are unaffected by the \B 's \uv\ light and
therefore can not originate from close to the \B. 2. There is no
significant evidence in the \x\ spectrum for Mn, which is known to
be extremely abundant on \mul. The methods used here are most
general and could be used to test whether other CP \B s are
intrinsic \x\ emitters, or not. The \x\ luminosity and variability
indicate that \mulb\ could be a \pmss, although it is impossible
to totally rule out a faint, late type cool star. By comparison
with grating observations of other \pmss s, we suspect the \x\
emission of \mulb\ is magnetically driven.

\acknowledgments This research was supported by SAO grant
GO3-4024X and by NASA/XMM grant NAG5-13266.  EB was supported by
the Yigal-Alon Fellowship, by ISF grant \#28/03, and by a grant
from the Asher Space Research Institute.

\clearpage
\bibliographystyle{apj}

\clearpage
\begin{figure}
\epsscale{.90}
\plotone{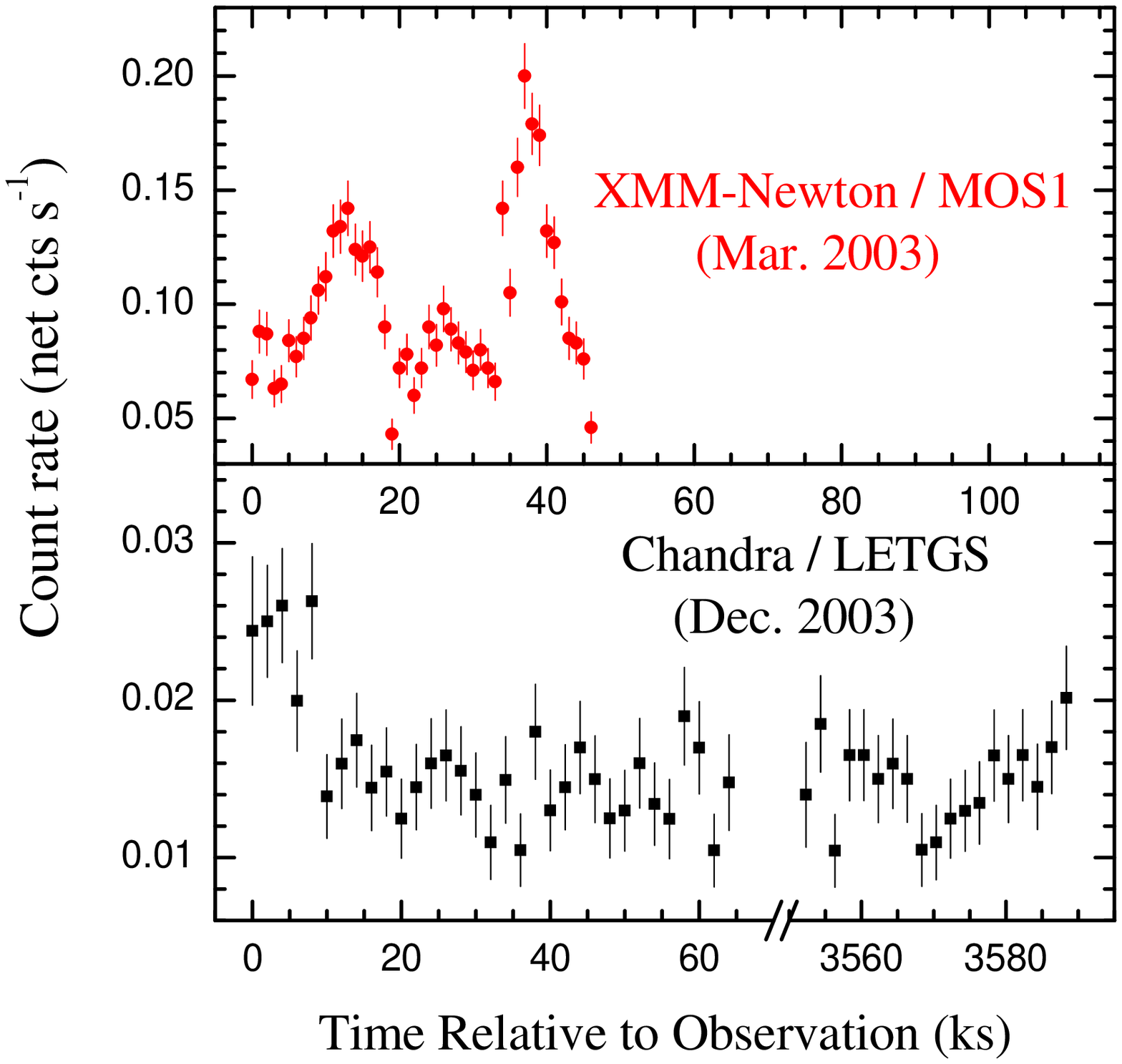}
\caption{Background subtracted
light curves of \mos 1 (top panel) and (not simultaneously)
zero-order \letgs\ (bottom) observations rebinned, respectively,
to 1 and 2~ks bins. Time is given relative to each observation.
Short duration flares can be seen. \label{fig1}}
\end{figure}

\clearpage
\begin{figure*}
\plotone{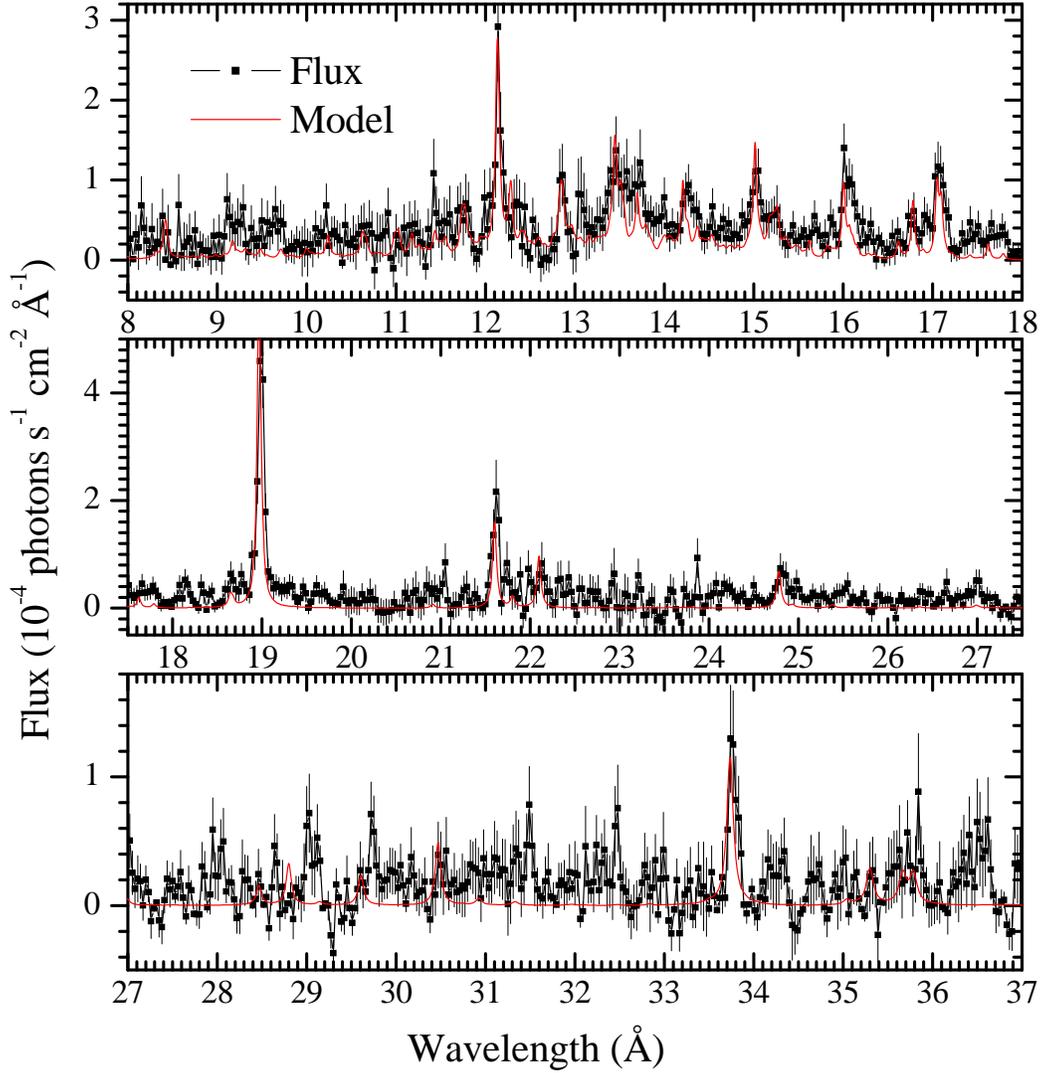}
\caption{Total 47~ks \rgs\ (1+2)
spectrum of \mulb, rebinned by a factor of three. The solid line
is the ion-by-ion model fitted to the data. The overall agreement
is good, but a few L-shell lines of mid-Z elements are still
missing in the model as seen by the discrepancies at the long
wavelength parts of the spectrum.
 \label{fig2}}
\end{figure*}

\clearpage
\begin{figure}
\epsscale{.90}
\plotone{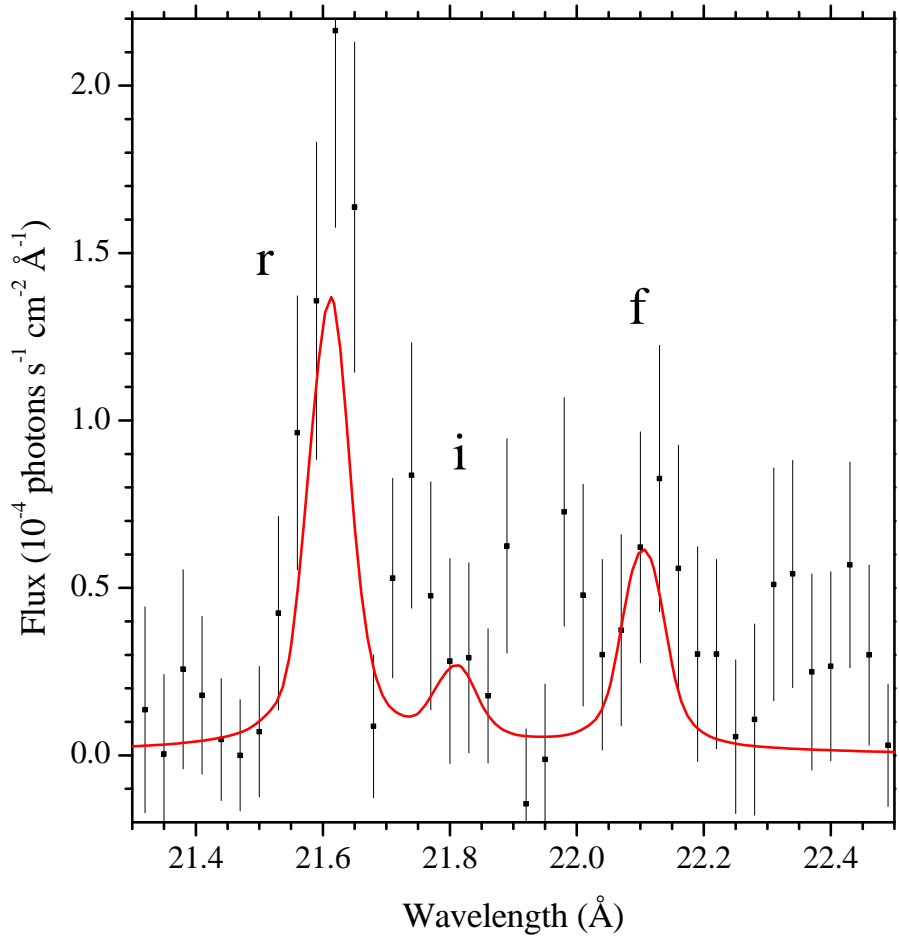}
\caption{\rgs\ spectrum of the
resonance (21.602~\AA), intercombination (21.802~\AA), and
forbidden (22.097~\AA) lines of He-like \ion{O}{7}. The solid line
represents the best-fit local 3-line model, only with the $i$-line
flux set to its upper limit still allowed by the data. (The noise
on both sides of the putative $i$ line is narrower than the
instrumental line spread function and therefore can not be
ascribed to the source.) \label{fig3}}
\end{figure}

\clearpage
\begin{figure}
\plotone{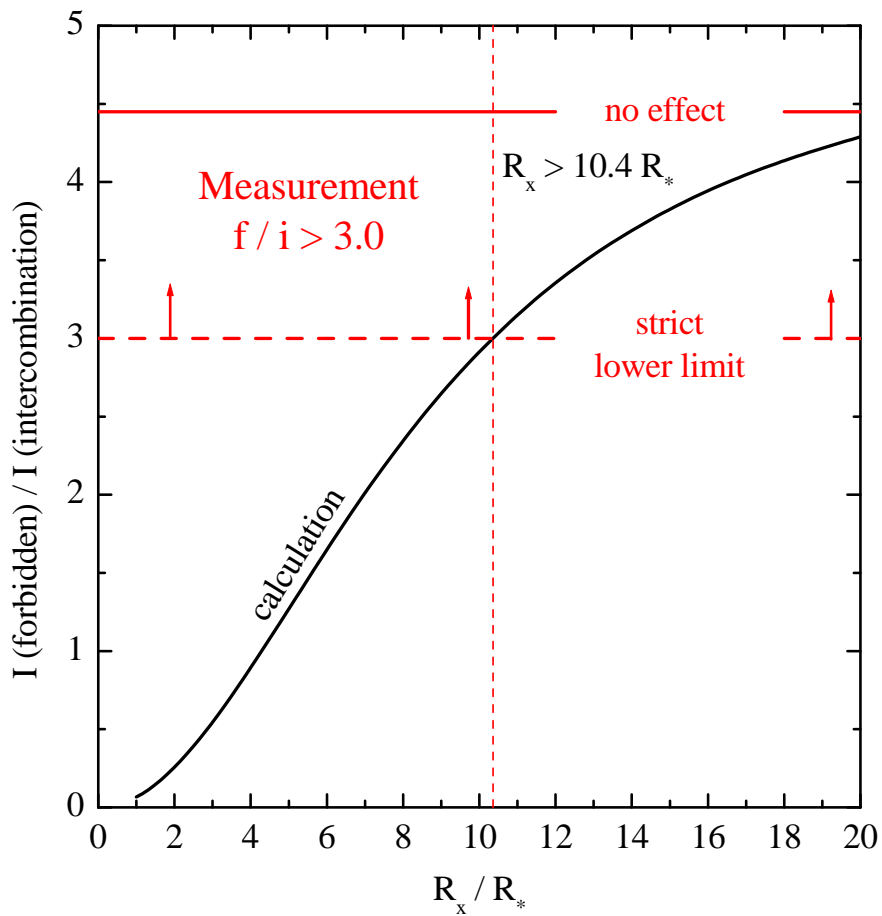}
\caption{The \ion{O}{7} {\it f /
i} calculated line intensity ratio as a function of the distance
of the \x\ source from \mul, assumed here to be a 12,400~K
blackbody source. The no-effect value ($\sim$4.4) and the strict
lower limit allowed by the current data (3.0) are indicated by
horizontal lines. The measurement requires that the \x\ source
(\mulb) be at least 10.4 stellar radii away from \mul.
\label{fig4}}
\end{figure}




\end{document}